\documentstyle[12pt]{article}
\begin{document}

\begin{titlepage}
\setcounter{page}{1}
\title{\bf On Solutions to the Nonlinear Phase Modification of the 
Schr\"{o}dinger Equation}
\author{Waldemar Puszkarz\thanks{Electronic address: puszkarz@cosm.sc.edu}
\\
\small{\it  Department of Physics and Astronomy,}
\\
\small{\it University of South Carolina,}
\\
\small{\it Columbia, SC 29208}}
\date{\small (May 13, 1999)}
\maketitle
\begin{abstract}
%\double
We present some physically interesting,  in general non-stationary, 
one-dimensional solutions to the nonlinear phase modification of the 
Schr\"{o}dinger equation proposed recently. The solutions  include a 
coherent state, a phase-modified Gaussian wave packet in the potential 
of harmonic oscillator whose strength varies in time, a free Gaussian 
soliton, and a similar soliton in the potential of harmonic oscillator 
comoving with the soliton. The last of these solutions implies that there 
exist an energy level in the spectrum of harmonic oscillator which is not 
predicted by the linear theory. The free solitonic solution can be considered 
a model for a particle aspect of the wave-particle duality embodied in the 
quantum theory. The physical size of this particle is naturally rendered 
equal to its Compton wavelength in the subrelativistic framework in which 
the self-energy of the soliton is assumed to be equal to its rest-mass energy. 
The solitonic solutions exist only for the negative coupling constant for 
which the Gaussian wave packets must be larger than some critical 
finite size if their energy is to be bounded, i.e., they cannot be point-like 
objects.

\vskip 0.5cm
\noindent
%PACS numbers: 11.80.Et, 12.10.Gq, 04.90.$+$e
\end{abstract}

\end{titlepage}

\section{Introduction}

%%\double

Recently we have presented the nonlinear phase modification of the
Schr\"{o}dinger equation \cite{Pusz1}. From the general scheme of the
modification we selected the two simplest models which guarantee that the
departure from the linear Schr\"{o}dinger equation is minimal in some
reasonable manner. One of the models turned out to have the same continuity
equation as the continuity equation of the Doebner-Goldin modification \cite
{Doeb} and we demonstrated that its Lagrangian leads to a particular variant
of this modification. The other model though constitutes a novel proposal
not investigated in the literature before. It is the purpose of this report
to present some physically interesting one-particle solutions\footnote{%
Multi-particle solutions are discussed in \cite{Pusz2}.} to this proposal
that we called the simplest minimal phase extension (SMPE) of the
Schr\"{o}dinger equation. Before doing so, let us briefly recall it.

In what follows, $R$ and $S$ denote the amplitude and the phase of the wave
function\footnote{%
We follow here the convention of \cite{Pusz1} that treats the phase as the
angle. In a more common convention $S$ has the dimensions of action and $%
\Psi =Rexp(iS/\hbar )$.} $\Psi =Rexp(iS)$, $V$ stands for a potential, and $%
C $ is the only constant of the modification that does not appear in linear
quantum mechanics. The discussed extension, similarly as the Schr\"{o}dinger
equation, is invariant under the Galilean group of transformations and the
space and time reflections. The Lagrangian density for the modification, 
\begin{equation}
-L(R,S)=\hbar R^{2}\frac{\partial S}{\partial t}+\frac{\hbar ^{2}}{2m}\left[
\left( \vec{\nabla}R\right) ^{2}+R^{2}\left( \vec{\nabla}S\right)
^{2}\right] +CR^{2}\left( \Delta S\right) ^{2}+R^{2}V,  \label{1}
\end{equation}
enables one to derive the energy functional, 
\begin{equation}
E=\int \,d^{3}x\,\left\{ \frac{\hbar ^{2}}{2m}\left[ \left( \vec{\nabla}%
R\right) ^{2}+R^{2}\left( \vec{\nabla}S\right) ^{2}\right] +CR^{2}\left(
\Delta S\right) ^{2}+VR^{2}\right\} ,  \label{2}
\end{equation}
a conserved quantity for the potentials that do not depend explicitly on
time which coincides with the quantum-mechanical energy defined as the
expectation value of the Hamiltonian for this modification \cite{Pusz1,
Pusz3}. The equations of motion for the modification read,

\begin{equation}
\hbar \frac{\partial R^{2}}{\partial t}+\frac{\hbar ^{2}}{m}\vec{\nabla}%
\cdot \left( R^{2}\vec{\nabla}S\right) -2C\Delta \left( R^{2}\Delta S\right)
=0,  \label{3}
\end{equation}
\begin{equation}
\frac{\hbar ^{2}}{m}\Delta R-2R\hbar \frac{\partial S}{\partial t}-2RV-\frac{%
\hbar ^{2}}{m}R\left( \vec{\nabla}S\right) ^{2}-2CR\left( \Delta S\right)
^{2}=0.  \label{4}
\end{equation}
As argued in \cite{Pusz1}, the most natural way to represent the nonlinear
coupling constant $C$ is as a product $\pm \hbar ^{2}L^{2}/m$, where $L$ is
some characteristic length to be thought of as the size of an extended
particle of mass $m$. This leads us to a nonlinear quantum theory of the
particle of mass $m$ and finite size $L$, but still leaves open the question
of sign of $C$. However, an alternative more traditional interpretation of $%
C $ as a universal coupling constant is also possible. In this approach $L$
is just a proxy for $C$ deprived of any additional physical meaning of its
own.

In general, the solutions to the modification do not possess the classical
limit in the sense of the Ehrenfest theorem. It is so because the Ehrenfest
relations for this modification contain some additional terms, 
\begin{equation}
m\frac{d}{dt}\left\langle \vec{r}\right\rangle =\left\langle \vec{p}%
\right\rangle +\frac{2Cm}{\hbar }\int \,d^{3}x\vec{r}\Delta \left( \Delta
SR^{2}\right) ,  \label{5}
\end{equation}
\begin{equation}
\frac{d}{dt}\left\langle \vec{p}\right\rangle =-\left\langle \vec{\nabla}%
V\right\rangle +C\int \,d^{3}x\left[ 2\vec{\nabla}S\Delta \left( \Delta
SR^{2}\right) -R^{2}\vec{\nabla}\left( \Delta S\right) ^{2}\right] .
\label{6}
\end{equation}
However, in the one-dimensional case, as long as $\Delta S=f(t)$, where $%
f(t) $ is an arbitrary function of time, these relations reduce to the
standard Ehrenfest relations.

For a system described by some Gaussian wave function an appropriate measure
of its physical size seems to be the width of its probability density. We
will take this measure as the definition of the physical size of the system.
As we will see, $L$ is related to the physical size of the system in some
situations that we will consider. Moreover, we will show that it is possible
to determine $L$ in in the so-called subrelativistic approach discussed in
connection with the solitonic solution. It is in this approach that $L$ can
be given a physical meaning only as a truly particle-dependent parameter of
the theory, the particle's attribute similarly as its mass or spin. We will
find that the physical size of the particle in this framework is that of its
Compton wavelength.

The stationary solutions to the linear Schr\"{o}dinger equation for which $%
S=-Et/\hbar +const$, where $E$ is the energy of a system, are also
stationary solutions to this modification. There may however exist other
stationary solutions as well. The purpose of the next two sections is to
present some non-stationary solutions that describe single systems in one
dimension, but, as we will see, the solitonic solution reduces to a
nontrivial stationary solution in the zero velocity limit.

\section{Wave Packet Solutions}

We will assume in this section that $\hbar =1$. Let us start from the
simplest solution that is also a solution to the linear Schr\"{o}dinger
equation. It is a coherent state for which 
\begin{equation}
R^{2}=\frac{1}{\sqrt{\pi }x_{0}}\exp \left[ {-\frac{\left( x-x_{0}\sqrt{2}%
\cos (\omega t-\delta )\right) ^{2}}{{x_{0}}^{2}}}\right]  \label{7}
\end{equation}
and 
\begin{equation}
S=-\left( \frac{\omega t}{2}-\frac{|\alpha |^{2}}{2}\sin 2\left( \omega
t-\delta \right) +\frac{\sqrt{2}|\alpha |x}{x_{0}}\sin \left( \omega
t-\delta \right) \right) .  \label{8}
\end{equation}
Since $\Delta S=0$, the coherent state being a solution to the linear
Schr\"{o}dinger equation in the potential of a simple harmonic oscillator, $%
V=m\omega ^{2}x^{2}/2$, represents a solution to equations (3) and (4) in
the very same potential. Here $x_{0}=1/\sqrt{m\omega }$, while $\alpha $ and 
$\delta $ are arbitrary constants, complex and real, respectively. The
physical size of this system is $x_{0}$, but since $\Delta S=0$, no relation
between the size in question and the characteristic size $L$ introduced by
the theory can be established.

The coherent state is an example of a wave packet. Another member of this
class, the ordinary Gaussian wave packet is not a solution to this
modification. Unlike the Gaussian packet, the coherent state does not spread
in time, but requires the potential of harmonic oscillator to support it.
Nevertheless, one can find another solution in this class that represents a
modified Gaussian wave packet whose amplitude is the same as that of the
ordinary Gaussian wave packet in the linear theory, 
\begin{equation}
R=R_{L}=\left[ \frac{mt_{0}}{\pi \left( t^{2}+t_{0}^{2}\right) }\right]
^{1/4}\exp \left[ -\frac{mt_{0}x^{2}}{2\left( t^{2}+t_{0}^{2}\right) }%
\right] .  \label{9}
\end{equation}
However, its phase is different from the phase of the ``linear'' packet, 
\begin{equation}
S_{L}=\frac{mtx^{2}}{2\left( t^{2}+t_{0}^{2}\right) }-\frac{1}{2}\arctan 
\frac{t}{t_{0}}.  \label{10}
\end{equation}
To ensure that this difference is minimal, we assume that the phase of the
modified packet has the form 
\begin{equation}
S=S_{L}+\frac{1}{2}f(t)x^{2}+h(t).  \label{11}
\end{equation}
The parameter $t_{0}$ is related to the average of the square of the
momentum of this system via $\left\langle p^{2}\right\rangle =m/2t_{0}$, so $%
t_{0}$ has to be positive. This is also necessary for the normalization of
the packet's wave function. Moreover, this constant determines the minimal
physical size of the system $L_{ph}^{2}(t=0)=$ $4t_{0}/m$. In principle,
this size could be arbitrarily small as the momentum of the packet can be
arbitrarily large. We will see however that for negative values of $C$ the
parameter $t_{0}$ must be larger than some finite value that depends on $C$.

Such a packet differs in the most minimal way from the Gaussian wave packet
of linear theory, but unlike the latter it may not exist without the support
of some external potential. We will now find the functions $f(t)$ and $h(t)$
and the potential $V(x,t)$ which is required to support this configuration.
Denoting for simplicity $\Delta S$ by $g(t)$, we find that the first
equation of the modification reduces to 
\begin{equation}
\frac{1}{m}\vec{\nabla}\left( xf(t)R^{2}-2Cmg(t)\vec{\nabla}R^{2}\right) =0.
\label{12}
\end{equation}
The solution of this equation is possible only if the expression in the
brackets is constant, but in order for the ratio $f(t)/g(t)$ to be a
function of time only this constant has to be zero. Consequently, one
obtains that 
\begin{equation}
\frac{f(t)}{2Cmg(t)}=-\frac{2mt_{0}}{t_{0}^{2}+t^{2}},  \label{13}
\end{equation}
and since $g(t)=\Delta S_{L}+$ $f(t)$, 
\begin{equation}
f(t)=-\frac{4Cm^{3}t_{0}t}{\left( t_{0}^{2}+t^{2}\right) \left(
t_{0}^{2}+t^{2}+4Cm^{2}t_{0}\right) }.  \label{14}
\end{equation}
The other equation of the modification will determine $h(t)$ and $V(x,t)$.
It boils down to 
\begin{equation}
\frac{1}{2}\dot{f}(t)x^{2}+\frac{f^{2}(t)x^{2}}{2m}+\frac{f(t)S_{L}^{\prime
}x}{m}+Cg^{2}(t)+\dot{h}(t)+V(x,t)=0,  \label{15}
\end{equation}
where overdots denote differentiation with respect to time and the prime
denotes differentiation with respect to $x$. Its solution requires that $%
V(x,t)=A(t)x^{2}$. One finds then that 
\begin{equation}
A(t)=\frac{2Cm^{3}t_{0}\left[ t^{4}-t_{0}^{3}(t_{0}+4Cm^{2})\right] }{\left(
t^{2}+t_{0}^{2}\right) ^{2}\left( t^{2}+t_{0}^{2}+4Cm^{2}t_{0}\right) ^{2}}
\label{16}
\end{equation}
and 
\begin{equation}
h(t)=-C\int dtg^{2}(t)=-Cm^{2}\int \frac{dtt^{2}}{\left(
t^{2}+t_{0}^{2}+4Cm^{2}t_{0}\right) ^{2}}.  \label{17}
\end{equation}
Calculating this integral gives 
\begin{equation}
h(t)=-\frac{Cm^{2}}{2}\left\{ \frac{1}{\sqrt{B^{2}(t_{0})}}\arctan \left( 
\frac{t}{\sqrt{B^{2}(t_{0})}}\right) +\frac{t}{t^{2}+B^{2}(t_{0})}\right\}
+const,  \label{18}
\end{equation}
when $B^{2}(t_{0})=t_{0}^{2}+4Cm^{2}t_{0}$ is non-negative, and 
\begin{equation}
h(t)=-\frac{Cm^{2}}{2}\left\{ \frac{t}{t^{2}-B^{2}(t_{0})}+\frac{1}{2a}\ln
\left| \frac{t+\sqrt{|B^{2}(t_{0})|}}{t-\sqrt{|B^{2}(t_{0})|}}\right|
\right\} +const  \label{19}
\end{equation}
otherwise. The energy of this configuration is time-dependent, 
\begin{equation}
E=\frac{t^{6}+3t_{0}^{2}t^{4}+t_{0}^{2}\left[ t_{0}^{2}+2t_{0}\left(
t_{0}+8Cm^{2}\right) \right] t^{2}+\left[ \left( t_{0}+4Cm^{2}\right)
^{2}+4Cm^{2}\left( t_{0}+4Cm^{2}\right) \right] t_{0}^{4}}{4t_{0}\left(
t^{2}+t_{0}^{2}\right) \left( t^{2}+t_{0}^{2}+4Cm^{2}t_{0}\right) ^{2}},
\label{20}
\end{equation}
and, as seen from this expression, it is asymptotically bounded by $%
E_{asymp}\equiv E(\left| t\right| \rightarrow \infty )=1/4t_{0}$. Therefore, 
$E_{asymp}=\left\langle p^{2}\right\rangle /2m$. The energy of the packet is
asymptotically conserved, but it changes locally in time due to the
time-dependent potential. Moreover, one observes that the energy of the
Gaussian scales as $1/|C|m^{2}$, which is precisely as anticipated in \cite
{Pusz1} based exclusively on dimensional arguments. A particularly simple
form of the formula for energy is obtained for the negative coupling
constant, $C=-|C|$, and $t_{0}=8|C|m^{2}$, 
\begin{equation}
E=\frac{t^{6}+3\left( 8Cm^{2}\right) ^{2}t^{4}+\left( 8Cm^{2}\right)
^{4}t^{2}}{16|C|m^{2}\left( t^{2}+\left( 8Cm^{2}\right) ^{2}\right) \left(
2t^{2}+\left( 8Cm^{2}\right) ^{2}\right) ^{2}}.  \label{21}
\end{equation}
We see that in this case, $E(t=0)=0$ and $E(t\neq 0)>0$.

What is the most interesting here is that the energy can become infinite for
negative values of $C$ unless $t_{0}>t_{0}^{cr,1}=4|C|m^{2}$. This critical
value of $t_{0}$ determines the lower bound on the minimal size of the
packet in question as discussed earlier. This bound cannot be attained.
Consequently, the lower bound for the minimal physical size of the packet is
related to the characteristic size as $L_{ph}^{lb,1}(t=0)=4L$. It is through
this relationship that $L$ could be, in principle, established
experimentally if the bound on the minimal physical size of the packet
proved to be somehow measurable. In the subrelativistic framework to be
discussed in the next section, $L=\lambda _{c}/4$, which leads to $%
L_{ph}^{lb,1}(t=0)=\lambda _{c}$. Let us also note that for the energy to be
non-negative, $t_{0}\geq t_{0}^{cr,2}=8|C|m^{2}$. Using $t_{0}^{cr,2}$ would
yield the higher lower bound on the minimal physical size of the packet
under study. In particular, in the subrelativistic framework, $%
L_{ph}^{lb,2}(t=0)=\sqrt{2}\lambda _{c}$. This bound is attainable. The
Gaussian wave packet under discussion does not alter the standard Ehrenfest
relations.

\section{Solitonic Solution}

We will now demonstrate that the modification discussed possesses a
solitonic solution. By the soliton we mean an object whose amplitude is well
localized and does not spread in time unlike that of ordinary Gaussian wave
packets. It should also be a solution to the nonlinear equations of motion,
i.e., we exclude the case of $\Delta S=0$. We will seek a solution that
resembles that of the Gaussian, but is not dispersive. Therefore, as an
Ansatz for the amplitude we take 
\begin{equation}
R(x,t)=N\exp \left[ -\frac{(x-vt)^{2}}{s^{2}}\right] ,  \label{22}
\end{equation}
where $v$ is the speed and $s$ is the half-width of the Gaussian amplitude
to be determined through the coupling constant $C$ and other fundamental
constants of the modification. The normalization constant $N=\left( 2/\pi
s^{2}\right) ^{1/{4}}$. We will seek the phase in the form 
\begin{equation}
S(x,t)=a(x-vt)^{2}+bvx+c(t),  \label{23}
\end{equation}
where $a$ and $b$ are certain constants and $c(t)$ is a function of time,
all of which need to be found from the equations of motion. Assuming that $%
V(x,t)=0$ and substituting (22) and (23) into (3) and (4) reveals that the
latter are satisfied provided 
\begin{equation}
b=m/\hbar ,s^{2}=-8mC/\hbar ^{2},s^{4}a^{2}=1,  \label{24}
\end{equation}
and 
\begin{equation}
2\hbar s^{4}m\frac{\partial c(t)}{\partial t}+2\hbar ^{2}s^{2}+\hbar
^{2}s^{4}b^{2}v^{2}+8Ca^{2}s^{4}m=0.  \label{25}
\end{equation}
We see that the coupling constant $C$ has to be negative, $C=-|C|$. From
(24) we now obtain that 
\begin{equation}
s^{2}=8m|C|/\hbar ^{2}=8L^{2}=8q\lambda _{c}^{2},  \label{26}
\end{equation}
\begin{equation}
a=\pm \hbar ^{2}/8m|C|=\pm \frac{1}{8L^{2}}=\pm \frac{1}{8q\lambda _{c}^{2}},
\label{27}
\end{equation}
where $q$ is the Compton quotient equal to $L^{2}/\lambda _{c}^{2}$ and $%
\lambda _{c}=\hbar /mc$ is the Compton wavelength of particle of mass $m$.
Combining (25-27) leads to 
\begin{equation}
c(t)=-\frac{1}{16}\left( \frac{\hbar ^{3}}{m^{2}|C|}+\frac{8mv^{2}}{\hbar }%
\right) t+const.  \label{28}
\end{equation}
The energy of the soliton is a function of its speed $v$, 
\begin{equation}
E\left( v\right) =E_{st}\left( L\right) +\frac{mv^{2}}{2},  \label{29}
\end{equation}
where 
\begin{equation}
E_{st}\left( L\right) =\frac{\hbar ^{2}}{16mL^{2}}=\frac{mc^{2}}{16q}
\label{30}
\end{equation}
is the stationary part of it. This part can become of the order of the rest
energy of the particle and even bigger for appropriately small $q$'s.
Nevertheless, as long as one remains outside the realm of special
relativity, the decay of particles due to energetic reasons is not an issue
and it is only the difference in the kinetic energy that matters and is
actually observed. This difference can be observed in the process of
changing the energy of the particle by slowing it down in some detector, in
particular by stopping it. In the latter case one detects that the change in
the particle's energy is $\Delta E=mv^{2}/2$. We also note the
characteristic scaling of energy being proportional to $\hbar ^{2}/mL^{2}$,
in agreement with what we anticipated in \cite{Pusz1}.

It is tempting to assume that the stationary energy term represents the rest
mass-energy of free particle, i.e., $E_{st}\left( L\right) =\hbar
^{2}/16mL^{2}=mc^{2}$. This determines the characteristic size $L$ of the
particle to be a quarter of its Compton wavelength. However, as seen from
(26), its physical size $L_{ph}=\sqrt{2}s=4L$ turns out to be equal to the
Compton wavelength itself. Implicit in this assumption is the fact that the
constant rest mass-energy term that one would obtain in the nonrelativistic
approximation is dropped from the scheme and replaced by the self-energy
term. The energy of this term gives rise to the rest mass-energy of the
particle. We call this approach subrelativistic. It leads to a model type of
particles whose physical size is precisely that of their Compton wavelength,
but in no way can it describe particles of any other size.\footnote{%
The requirement that the size of the quantum particle should be that of its
Compton wavelength has recently been used as a postulate to build a model of
nonlinear quantum mechanics of extended objects from first principles \cite
{Sas}.} It seems that the most appropriate way to interpret these solitons
is as the fundamental particular constituents of quantum realm complementary
to waves. Quantons, as we choose to call them, would then be the unique
realization of the particle aspect of the wave-particle duality of quantum
mechanics.

This all seems to be too easy so one can suspect some trick here. The trick
is that out of three constants $\hbar $, $m$, and $L$ the last two having
dimensions of kg and meter, respectively, it is always possible to form a
quantity of the dimensions of energy, $\hbar ^{2}/mL^{2}$, and if this
quantity is to be of the order of the rest mass-energy of the particle then $%
L$ should be of the order of its Compton wavelength $\lambda _{c}$. However,
it is not necessarily as easy as this simple reasoning may suggest. First of
all, this dimensional trick does not imply that the physical size of the
quanton is to be precisely equal to its Compton wavelength. The fact that it
is so is thus rather remarkable. Secondly, and even more importantly, if a
good joke is not to be repeated too often ours is a good one indeed for it
cannot be repeated neither in the Doebner-Goldin \cite{Doeb} nor in the
Bia\l ynicki-Birula modification \cite{Bial1}, although for two different
reasons. In the former, the dimensions of its nonlinear parameters do not
allow to make any new dimensional quantities beyond those that can be made
up of $\hbar $ and $m$ and those two constants are not enough for our task.
In the latter, the nonlinear parameter $\varepsilon $ has the dimensions of
energy and the dimensional analysis of the problem implies that the
characteristic size of an object of such energy is inversely proportional to
the square root of it. Indeed, the soliton of this modification has the
radius $\hbar /\sqrt{2m\varepsilon }$. Experimentally established \cite{Gahl}%
, the upper bound on the value of this parameter is so small that it implies
the existence of objects of macroscopic size and thus easy, in principle, to
observe. Nevertheless, they have not been empirically confirmed.

One can however consider the self-energy independent of the rest-mass
energy. The rest mass-energy would then constitute a separate part of the
total energy or it could be eliminated from the considerations in a
completely non-relativistic framework. None of these approaches is more
advantegeous than the other, both are just models. The first of them
attempts to model the rest mass-energy of the quanton and thus its inertia
by means of the self-interaction term, the other approach is devoid of such
a goal and treats the rest mass-energy as given and inconsequential.

Let us now present the subrelativistic formulation in a more mathematical
manner. As a subrelativistic Hamiltonian $H_{sub}$ of the free
Schr\"{o}dinger equation we define the Hamiltonian whose expectation value
on its solutions is 
\begin{equation}
<H_{sub}>=E=\frac{mv^{2}}{2}+mc^{2}.  \label{31}
\end{equation}
It is through this equation that the quantum and classical world make
contact. However, this equation by no means fixes the form of
subrelativistic Hamiltonian. It is easy to find such a Hamiltonian in linear
quantum mechanics. In fact, it is unique and it differs from the Hamiltonian
of the free Schr\"{o}dinger equation $H_{Sch}$ by the rest energy term. In
other words, it is 
\begin{equation}
H_{sub}^{L}=H_{Sch}+H_{rest}^{L}=H_{Sch}+mc^{2}.  \label{32}
\end{equation}
The solution to the free linear subrelativistic Schr\"{o}dinger equation is,
similarly as in the nonrelativistic case, a plane wave $\Psi =\exp
(iS_{plane})$, but with a slightly different phase, 
\begin{equation}
S_{plane}=\frac{mvx}{\hbar }-\left( E-mc^{2}\right) t.  \label{33}
\end{equation}
In general, the rest energy term is of no relevance in the linear
formulation of the subrelativistic Schr\"{o}dinger equation for it can be
absorbed in the phase of the quantum system without any further
consequences. In nonlinear quantum mechanics, things can be very different.
Again, we expect that the following decomposition 
\begin{equation}
H_{sub}^{NL}=H_{Sch}+H_{rest}^{NL}  \label{34}
\end{equation}
will lead to the dispersion relation (31). Now, however, the choice of the
rest energy Hamiltonian $H_{rest}^{NL}$ is not unique. It seems that the
most reasonable and minimal in some sense way to enhance this uniqueness is
to stipulate that a free nonlinear subrelativistic Schr\"{o}dinger equation
has a solitonic solution which satisfies (31). As argued above, the
Doebner-Goldin modification is unable to produce (31) for its nonlinear
solutions due to dimensional reasons. The Bia\l ynicki-Birula and Mycielski
modification can be thought of as a subrelativistic nonlinear extension of
the Schr\"{o}dinger equation only for a sufficiently light particle due to
the experimental smallness of $\varepsilon $. It is not out of the question
that the SMPE is the only nonlinear modification of the Schr\"{o}dinger
equation that can be considered a nonlinear subrelativistic Schr\"{o}dinger
equation which has a solitonic solution fulfilling (31) and which, in
addition, entails the unique value for the physical size of the soliton. Let
us note that the plane wave that is the solution of the subrelativistic
linear Schr\"{o}dinger equation is also a solution to the modification in
question. Nevertheless, it should be noted that the coupling constant of
this modification is no longer universal in this approach for it is
determined by other parameters of the theory to the effect that $%
C_{sub}=-\hbar ^{4}/16m^{3}c^{2}$. Consequently, if $C$ is ever
experimentally found to be independent of the mass of the particle, i.e., $C$
is indeed a truly universal constant and not a product of $\hbar ^{2}$, $m$,
and $c$, then the discussed approach is viable only for one mass and thus is
much less appealing. For this reason, it is more appropriate in this case to
use $L$ rather than $C_{sub}$, for the former, being the characteristic size
of the particle represents its attribute and therefore cannot be thought of
as a universal constant.

A similar solitonic solution exists also in the following time-dependent
potential of harmonic oscillator, 
\begin{equation}
V(x,t)=k(x-vt)^{2},  \label{35}
\end{equation}
for any negative value of the coupling constant $C$. The amplitude and phase
of the soliton are assumed to be the same as before, i.e., given by (22) and
(23). The parameter $b$ is determined by (24) and the half-width of the
soliton by (26), and so none of them is affected by the potential. Moreover $%
c(t)$ is also determined by (25), except that now $a$ satisfies the equation 
\begin{equation}
a^{2}=\frac{1}{s^{4}}-\frac{km}{2\hbar ^{2}}=\frac{1}{64L^{4}}-\frac{km}{%
2\hbar ^{2}},  \label{36}
\end{equation}
which implies that the strength of the potential cannot be greater than $%
k_{crit}=\hbar ^{2}/32mL^{4}=mc^{2}/32qL^{2}$. Choosing the standard form of 
$k$, $k=m\omega ^{2}/2$, we obtain that for a fixed $L$, $\omega \leq $ $%
\omega _{crit}=\hbar /4mL^{2}$. For a given $\omega $, $L\leq L_{\max }=%
\sqrt{\hbar /4m\omega }$. The energy of this configuration is 
\begin{equation}
E\left( v,L;d\right) =E_{st}\left( L;k\right) +\frac{mv^{2}}{2},  \label{37}
\end{equation}
where 
\begin{equation}
E_{st}\left( L;d\right) =\frac{\hbar ^{2}}{16mL^{2}}+2kL^{2}=\frac{mc^{2}}{%
16q}+2qk\lambda _{c}^{2}  \label{38}
\end{equation}
represents the stationary part of it. We note that it is only the phase and
the energy of the particle that depend on the potential. The average
position $\left\langle x\right\rangle =vt$ and momentum $\left\langle
p\right\rangle =mv$ are the same for both of these solitonic solutions. In
the case when $v=0$, each of these solutions reduces to a stationary
solution of energy $E_{st}\left( L\right) $ and $E_{st}\left( L;k\right) $,
respectively.

For a given $L$, the maximum stationary energy (38) equals $E_{st}\left(
L;k_{crit}\right) =\hbar ^{2}/8mL^{2}=mc^{2}/8q$. However, as a function of $%
L$, $E_{st}\left( L;k\right) $ does not have a maximum, but a minimum. This
minimum is attained for $L=L_{\max }$ and it is equal to the ground state
energy of the harmonic oscillator, $E=\hbar \omega /2$. Since for $L_{\max }$%
, $a=0$ and $\Delta S=0$, we see that this state corresponds to the ground
state of linear theory.

For $L<L_{\max }$, there exist two nodeless wave functions of which one
corresponds to the ground state of linear theory ($\Delta S=0$). The state
described by the other wave function has the energy, 
\begin{equation}
E_{st}\left( \omega \right) =\frac{\hbar \omega }{4}\left( \frac{1}{Q_{h}}%
+Q_{h}\right) =\frac{\hbar \omega _{crit}}{4}\left( 1+\frac{\omega ^{2}}{%
\omega _{crit}^{2}}\right) =mL^{2}\left( \frac{\hbar ^{2}}{16m^{2}L^{4}}%
+\omega ^{2}\right) ,  \label{39}
\end{equation}
where $Q_{h}=\left( L/L_{\max }\right) ^{2}=4L^{2}m\omega /\hbar =4q\hbar
\omega /mc^{2}=\omega /\omega _{crit}$. Therefore, 
\begin{equation}
\Delta E_{new}=E_{st}-E_{g}=\frac{\hbar \omega }{4}\left( \frac{1}{Q_{h}}%
+Q_{h}-2\right) =\frac{\hbar \omega _{crit}}{4}\left( 1-\frac{\omega }{%
\omega _{crit}}\right) ^{2}  \label{40}
\end{equation}
represents a new line in the spectrum of harmonic oscillator not predictable
by the linear theory. In terms of the separation between consecutive energy
levels $E_{con}$ in the spectrum of linear theory and the frequency ratio $%
\eta =\omega /\omega _{crit}$, ($\eta \leq 1$) 
\begin{equation}
\frac{\Delta E_{new}}{E_{con}}=\frac{1}{4}\left( \frac{1}{Q_{h}}%
+Q_{h}-2\right) =\frac{1}{4\eta }\left( 1-\eta \right) ^{2}.  \label{41}
\end{equation}
In principle, it is easy to verify the existence of the new line. One should
start observing the spectrum of harmonic oscillator right below $\omega
_{crit}$. It is at this critical frequency that the new line splits off of
the ground state and as we keep lowering the frequency, it moves towards the
first excited state of linear theory. At $\eta =1/4$ it is approximately
half-way there. The critical frequency $\omega _{crit}$ expressed in Hz is
approximately 
\begin{equation}
\omega _{crit}=3\times 10^{19}\frac{m}{q_{e}m_{e}},  \label{42}
\end{equation}
where $m_{e}$ is the mass of electron and $q_{e}$ the Compton quotient of a
particle with respect to the Compton wavelength of the electron. Even for
the lightest stable particle, the electron, this is well above the top range
of frequency of gamma rays of the order of $10^{7}$ Hz, which seems to make
impossible to carry out this type of experiment. We should note that if $%
\omega \ll \omega _{c}$, which is a much more accessible regime, the new
level has to be sought among highly excited states of harmonic oscillator as
seen from (41). This may not necessarily be feasible either.

It is reasonable to expect that $L$ is of the order of $\lambda _{c}$. In
the subrelativistic framework, it is $L=\lambda _{c}/4$ that should be
chosen. This yields the formula 
\begin{equation}
E_{st}=mc^{2}+\frac{m\lambda _{c}^{2}\omega ^{2}}{8}  \label{43}
\end{equation}
valid for $\omega \leq \omega _{crit}=4\hbar /m\lambda _{c}^{2}=4m/\hbar $.
However, it seems more appropriate to impose the condition $m\lambda
_{c}^{2}\omega ^{2}<8$, so that the particle creation-annihilation does not
occur. This defines the subrelativistic frequency regime to be $\omega
<\omega _{creat}=2\sqrt{2}m/\hbar $. What we obtained is a hard core
particle regime with the physical size of the oscillator depending only on
universal constants and equal to its Compton wavelength, in contrast to a
``soft'' core type of oscillator of linear theory whose size can be modified
by changing its frequency. If this modification describes reality then we
should be able to observe that one of the energy levels in the spectrum of
the harmonic oscillator depends quadratically on $\omega $.

Solitonic solutions occur also in other nonlinear modifications of the
Schr\"{o}dinger equation, as, for instance, in the modification of Bia\l
ynicki-Birula and Mycielski \cite{Bial1, Bial2, Ofic, Min} and in the
Doebner-Goldin type of modifications \cite{Natt1, Natt2, Capa1, Capa2}. It
should be pointed out that the solitons presented in this paper exist for
arbitrary values of the (negative) coupling constant which is not always the
case in other nonlinear modifications where for this to happen some
threshold value of nonlinear parameter(s) must be exceeded.

\section{Conclusions}

We have presented four non-stationary one-dimensional solutions to the
simplest minimal phase extension of the Schr\"{o}dinger equation introduced
in \cite{Pusz1}. The simplest of them, being also a solution to the linear
Schr\"{o}dinger equation, represents a coherent state which is a particular
form of a wave packet. Its existence requires the potential of harmonic
oscillator. Similar in nature is the second solution, the modified Gaussian
packet whose amplitude is identical with the amplitude of the ``linear''
Gaussian wave packet, its phase being slightly different but having the same
spatial shape as the phase of the ordinary Gaussian packet. This solution
exists in the potential of harmonic oscillator with a time-dependent
strength. The wave packet in question is dispersive, which is not the case
for the coherent state and the other two solutions, the free Gaussian
soliton and a similar soliton in the potential of harmonic oscillator
travelling with the velocity of the soliton. These two objects are
characteristic of nonlinear structures. All of these solutions have the
standard Ehrenfest limit.

For the existence of the solitons it is necessary that $C<0$. The other
solutions exist for any value of the coupling constant, but it is only for
the negative $C$ that the Gaussian packet seems to corraborate our
hypothesis that the theory discussed describes extended particles. Indeed,
if the coupling constant is negative, the minimal physical size of the
packets must be larger than some finite value for otherwise they would
develop infinite energy at some point. This squares quite nicely with the
idea of extended, i.e., not point-like particles.

The most physically interesting of the solutions presented is the free
solitonic solution. It is conceivable that this solution can serve as a
particle representation of the wave-particle duality embodied in quantum
mechanics. The standard quantum theory despite many successful years of
development has not been able to provide an acceptable physical realization
of this duality as only the wave aspect of the duality in question has been
incorporated in the mathematical structure of the theory. The wave packets
cannot serve as good models of particles for they spread in time, suggesting
that there exist macroscopically extended quantum objects contrary to the
empirical evidence in this matter. The fact that these packets are not free
solutions to the SMPE can thus be viewed as a partial boon to the theory,
even if the theory implies that it is possible to create similar wave
packets if an appropriate time-dependent potential is applied. Other notable
modifications of the Schr\"{o}dinger equation also contain wave packet
solutions for time-dependent potentials.

A good mathematical model of the particle should represent an object that is
well localized and non-dispersive. The free soliton presented in this paper
meets these requirements. What is specially attractive about it is that it
is a particle solution to the modification that does not alter well verified
properties of the quantum world established by pure wave mechanics such as,
for instance, the atomic structure. This solution seems to be particularly
relevant in the context of de Broglie-Bohm formulation of quantum mechanics 
\cite{Bohm}. It is this formulation that puts a considerable emphasis on the
particle aspect of the wave-particle duality. Whereas in the Copenhagen
interpretation of this theory it is either the wave or the particle, and the
particle can be viewed as the result of interference of waves, in the
approach pioneered by de Broglie it is both the wave and the particle. In
this picture, the waves are always associated with particles and serve as
guides for them according to the original de Broglie idea of pilot waves 
\cite{Brog}. Needless to say that without a particle solution to the
equations of motion, this picture is rather incomplete. The free particle
solution of our modification can coexist with any solution of linear wave
mechanics in the sense that they can be part of a bigger system described by
a factorizable wave function without violating separability. This is indeed
a perfect marriage of wave and particle in that they always remain separated.

Other nonlinear modifications also contain particle-like solutions that
might fulfill the dream of de Broglie. In \cite{Vig}, in a model
specifically designed for this purpose the existence of a class of possible
solutions of particle-like properties is demonstrated. However, these are
solutions to approximate nonlinear equations. In the Bia\l ynicki-Birula and
Mycielski modification, the width of free Gaussian soliton is $\hbar /\sqrt{%
2m\varepsilon }$, where $\varepsilon $ is the only nonlinear physically
significant parameter of the theory. Since the current upper bound on this
parameter is \cite{Gahl} $3.3\times 10^{-15}$ eV, it implies that the size
of gausson of the electron mass is of the order of $3$ mm which is a
macroscopic value! Such solitons would be easy to observe, but so far they
have somehow managed to escape our attention. It is thus likely that they
simply do not exist. A remarkable class of new type of solitons,
finite-length solitons,\footnote{%
This type of solitonic solutions are known as compactons in the broader
literature.} have been recently discovered in the Doebner-Goldin type of
modifications \cite{Capa2}. As observed in \cite{Capa2}, `they realize the
``dream of De Broglie,'' in the sense that they permit to identify a quantum
particle with a non-spreading wave-packet of finite length travelling with a
constant velocity in the free space.' However, the length in question
depends on the speed and the frequency of the soliton, and in some cases the
smaller these are the bigger the length of the soliton. In particular
circumstances nothing can prevent this length from becoming arbitrarily
large, and so if these objects are to resemble microscopic quantum particles
some additional physically justifiable assumptions are necessary. The
Doebner-Goldin modification itself does not seem to provide any insight on
how to handle this problem, in part because the physical meaning of its
parameters is not well elucidated.

On the other hand, the width of the solitons found in this paper which is a
measure of their localization is of the order of the characteristic length
of the modification, the length of the extended particle-system which this
theory can be thought of describing. It seems rather unlikely that one can
find a soliton of reasonably small size for an arbitrary value of a
nonlinear coupling constant that would be a physically sound model of
quantum particle in a theory which does not involve implicitly or explicitly
a parameter proportional to some characteristic length or its power. The
examples presented in the preceding paragraph were intended to illustrate
precisely this point.

The stationary soliton solution in the potential of harmonic oscillator
implies that there exist an energy level in the spectrum of harmonic
oscillator not predictable by the linear theory. The energy of this level
depends on the characteristic size of the oscillator that is limited by a
certain critical value $L_{\max }$ which corresponds to the linear theory.
It is for this value that the level in question coincides with the ground
state of the harmonic oscillator in linear quantum mechanics and attains its
minimum. The solution that corresponds to the nonlinear theory must thus be
more compact than the solution of linear theory. Indeed, unlike the
discussed solution of linear quantum mechanics with a ``soft'' size of the
oscillator that can be modified by changing its frequency, the nonlinear
solution describes a hard core particle regime with the physical size of the
oscillator depending only on universal constants. If this modification
describes reality then we should be able to observe that one of the energy
levels in the spectrum of the harmonic oscillator depends quadratically on $%
\omega $.

Finally, let us note that a particularly fitting approach to the SMPE as the
theory of extended particles is the approach that we term subrelativistic.
This approach introduces the speed of light $c$, as in the rest mass-energy
of a system, but the framework of special relativity is not needed; the
Galilean transformation is the symmetry of the theory. Being
nonrelativistic, the Schr\"{o}dinger equation provides only a limited
description of physical phenomena. One can derive it from the Klein-Gordon
equation in the limit in which the Compton wavelength is much smaller than
de Broglie's wavelength of quantum particle. Yet, the Klein-Gordon equation
cannot be used as an equation for a generic relativistic spinless quantum
system due to the problem of negative probabilities. It is tempting to
extend the limits of the Schr\"{o}dinger equation to the domain between the
completely nonrelativistic and relativistic world, to the subrelativistic
realm. Subrelativistic phenomena are not necessarily of only speculative
character. They may arise due to certain pecularities of nonrelativistic
quantum mechanics that, as argued in \cite{Dieks}, does not in all respects
behave as a fully Galilean invariant theory as one would expect it in the
nonrelativistic limit. The difference is empirically significant, as
illustrated by the Sagnac effect \cite{Anand1}, and is due to the fact that
the ``quantum'' Galilei group is not identical with its classical
counterpart \cite{Levy} for the former bears the remnants of its
relativistic origin. Therefore, the nonrelativistic quantum-mechanical
description is sometimes inevitably subrelativistic as ultimately based on a
broader group of symmetry than the classical Galilei group. Other effects of
this kind that justify the subrelativistic approach may, in principle, be
possible too. A particularly important instance of such an effect is
provided by the spin-orbit coupling.

It is within the subrelativistic approach that one can uniquely determine
the physical size of the free particle which turns out to be equal to its
Compton wavelength, a most reasonable size for a quantum particle. However,
the approach in question makes sense only if the nonlinear parameter of the
theory is particle-dependent, i.e., this parameter, such as $L$, has to be
the particle's attribute in the same manner as its mass and not a universal
constant.

\section*{Acknowledgments}

I would like to thank Professor Pawe{\l } O. Mazur for bringing my attention
to the work of Professor Staruszkiewicz that started my interest in
nonlinear modifications of the Schr\"{o}dinger equation. I am also grateful
to Professor Andrzej Staruszkiewicz for the critical reading of the
preliminary version of this paper, his comments and a discussion, and to
Kurt Ko{\l }tko for his interest in this work. A correspondence with Dr.
Marek Czachor whose comments helped in a better presentation of some of the
ideas of this paper is particularly acknowledged. This work was partially
supported by the NSF grant No. 13020 F167 and the ONR grant R\&T No. 3124141.

\end{document}